\documentclass[%
reprint,
amsmath,amssymb,
aps,
]{revtex4-2}
\usepackage{mhchem}%
\usepackage{graphicx}%
\usepackage{dcolumn}
\usepackage{bm}
\usepackage{caption}
\usepackage{subfigure}
\usepackage{multirow}
\usepackage{booktabs}
\usepackage[colorlinks=true, letterpaper=ture, pdfstartview=FitV, linkcolor=blue, citecolor=blue, urlcolor=blue]{hyperref}
\begin{document}

\preprint{APS/123-QED}

\title{Zero-Point Quantum Diffusion of Proton in Hydrogen-rich Superconductor $\ce{LaH10}$ from First-principles}

\author{Xuejian Qin$^{1,2}$}
\thanks{These authors contributed equally to this work.}
\author{Hongyu Wu$^{1,2}$}
\thanks{These authors contributed equally to this work.}
\author{Guoyong Shi$^{3}$}
\author{Chao Zhang$^{3}$}
\author{Peiheng Jiang$^{1}$}
\author{Zhicheng Zhong$^{1,4}$}
\thanks{zhong@nimte.ac.cn}

\affiliation{\\$^1$CAS Key Laboratory of Magnetic Materials and Devices $\&$ Zhejiang Province Key Laboratory of Magnetic Materials and Application
Technology, Ningbo Institute of Materials Technology and Engineering, Chinese Academy of Sciences, Ningbo 315201, China}

\affiliation{$^2$College of Materials Science and Opto-Electronic Technology, University of Chinese Academy of Sciences, Beijing 100049, China}

\affiliation{$^3$Department of Physics, Yantai University, Yantai, 264005, China}

\affiliation{$^4$China Center of Materials Science and Optoelectronics Engineering, \\
University of Chinese Academy of Sciences, Beijing 100049, China}


\date{\today}

\begin{abstract}
$\ce{LaH10}$, as a member of hydrogen-rich superconductors, has a superconducting critical temperature of 250 K at high pressures, which exhibits the possibility of solving the long-term goal of room temperature superconductivity. Considering the extreme pressure and low mass of hydrogen, the nuclear quantum effects in $\ce{LaH10}$ should be significant and have an impact on its various physical properties. Here, we adopt the method combines deep-potential (DP) and quantum thermal bath (QTB), which was verified to be able to account for quantum effects in high-accuracy large-scale molecular dynamics simulations. Our method can actually reproduce pressure-temperature phase diagrams of $\ce{LaH10}$ consistent with experimental and theoretical results. After incorporating quantum effects, the quantum fluctuation driven diffusion of proton is found even in the absence of thermal fluctuation near 0 K. The high mobility of proton is found to be compared to liquid, yet the structure of $\ce{LaH10}$ is still rigid. These results would greatly enrich our vision to study quantum behavior of hydrogen-rich superconductors.
\end{abstract}

\maketitle


\section{Introduction}
Recently, the development of high-$T_c$ hydrogen-rich superconductors has attracted a lot of attention\cite{Drozdov2015,Semenok2020,Ma2022,FloresLivas2020}, because chemical precompression\cite{Gilman1971,Ashcroft2004} has emerged as an effective approach to reduce pressure in the preparation of hydrogen-rich superconductors. Among these superconductors, $\ce{LaH10}$ has demonstrated a high-$T_c$ of around 250K at 170-200 GPa, and has been suggested as a potential candidate for room-temperature superconductivity \cite{Somayazulu2019,Drozdov2019,Sun2021}.

The nuclear quantum effects of the zero-point energy and fluctuations are strong and play a key role in $\ce{LaH10}$. The zero point energy $\frac{1}{2}\hbar\omega$ of a specific system can be approximated by $\omega = \sqrt{K/M}$, where $\omega$ and $K$ are the effective frequency and the spring constant, $M$ is the mass of atom. Due to the special environment under high pressure, the effective spring coefficient of $\ce{LaH10}$ is remarkably high. Besides, hydrogen has a significntly low mass. The combination of these two factors leads to the abnormal strong nuclear quantum effects in $\ce{LaH10}$. Nuclear quantum effects have been shown to be crucial in stabilizing superconductivity favored crystalline structure of $\ce{LaH10}$ \cite{Errea2020,Watanabe2022}. 

Another impact of nuclear quantum effects is the motion of proton. Since the nuclear quantum effect is strong in $\ce{LaH10}$ and atom mass of hydrogen is low, the zero-point fluctuations of atomic position should be strong and may be comparable to its near neighbor distance. Similar to the superionic behavior exhibited by water under high pressures, where oxygen atoms retain localization while protons diffuse through the lattice\cite{Kapil2022}. Similar behaviors exist in quantum crystals\cite{Cazorla2017}, which possess some unique physical properties such as high mobility of atoms, thus unique behaviors may also be found in hydrogen-rich superconductor $\ce{LaH10}$.

However, achieving large scale, first-principles simulation with quantum effects of zero-point energy is full of challenge. For example, density-functional-theory (DFT) is not able to directly incorporate quantum effects of zero-point energy and is hard to simulate over thousands atoms. Deep-potential (DP) + quantum thermal bath (QTB) strategy which is capable of large-scale atomistic dynamic simulation with density-functional-theory (DFT) precision. This strategy is \textit{ab initio}, efficient and proven to be able to correctly describe the complex phenomena caused by quantum effects in our previous work\cite{Wu2022}. DP is a machine-learning method which can produce accurate force fields of molecular dynamics (MD) by sampling the results of the DFT\cite{Zhang2018,Zhang2018a}. QTB is a method that preserves the features of quantum statistics in MD\cite{Dammak2009}. Its core idea is rewriting the fundamental Newtonian equations in MD into Langevin-like equations with colored noise modified with zero point energy. DP+QTB requires a comparable computation cost to classical MD, which can easily simulate a few millions of atoms. Thus we can provide results with large space and time scale from which we can obtain sufficient physical information.

In this work, we employ deep-potential (DP)+ quantum thermal bath (QTB) method to investigate the nuclear quantum effects of high-pressure $\ce{LaH10}$. Our pressure-temperature ($P$-$T$) phase diagram of $\ce{LaH10}$ results show that the quantum effects make the cubic phase more stable, consistent with previous findings. More strikingly, we find that in the absence of thermal fluctuations near 0 K, the quantum zero point fluctuation will induce the liquid-like diffusion of proton, while the La framework remains stable.

\begin{figure*}[htbp]
    \centering
    \includegraphics[width=0.85\textwidth]{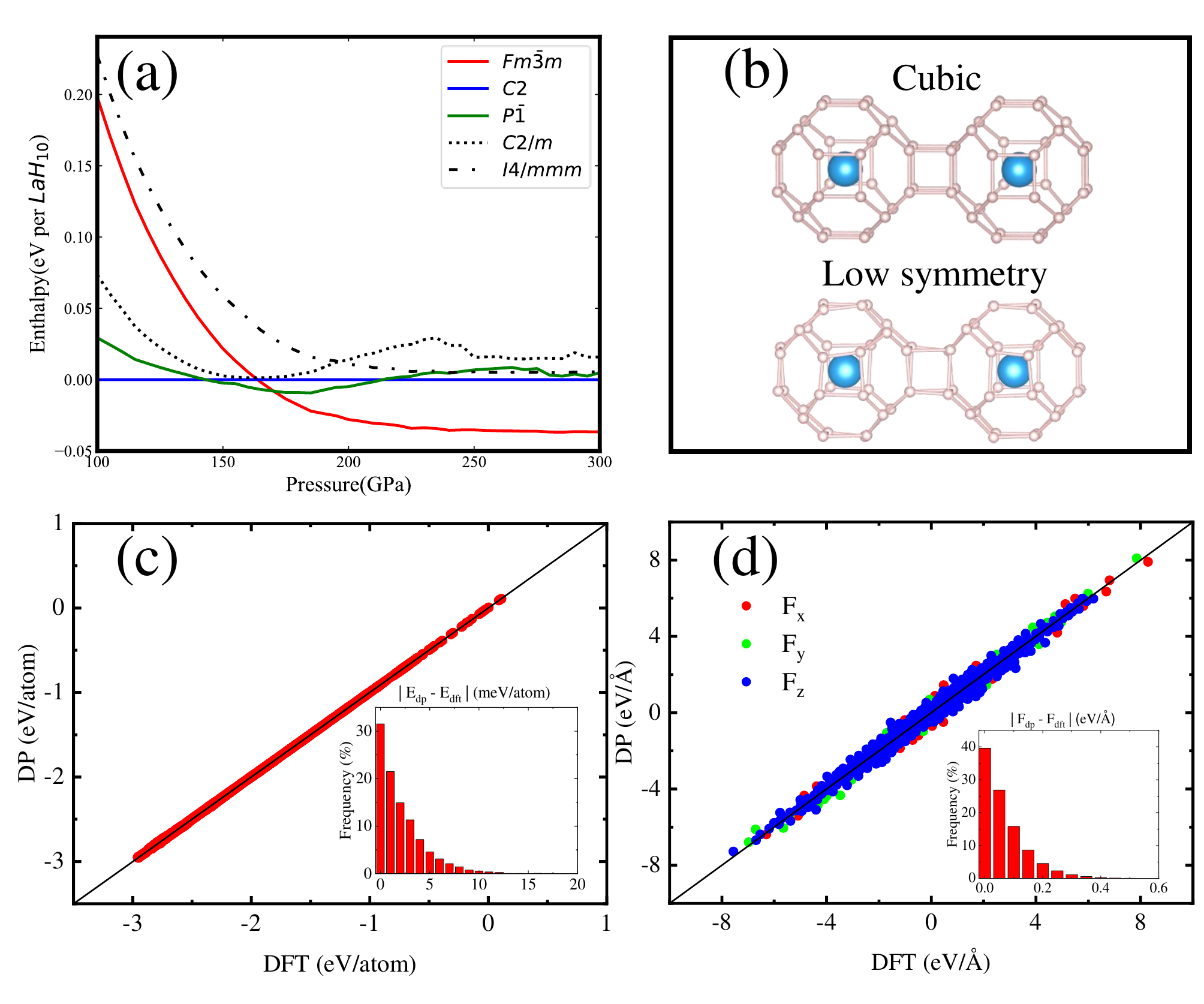}
    \caption{(a) Enthalpy (using the values of the $C2$ group structure as  reference) as a function of pressures for different structures searched by USPEX . (b) Schematic of the high symmetry $Fm\bar{3}m$ phase (above), and the low symmetry $C2$ and $P\bar{1}$ phase (below). Comparison of (c) energies and (d) atomic forces calculated using the DP model and DFT, respectively.}
    \label{fig:method_part}
\end{figure*}

\section{Computational Method}

\subsection{Deep Potential of $\ce{LaH10}$}

DP is a machine-learning-based method that aims to provide highly applicable and high-accuracy interatomic interaction potentials\cite{Zhang2018}. The core idea of DP model is to fit the energy and force of different configurations of certain material through a deep neural network, and its learning samples are a large number of structural configurations ( usually these configurations have no more than 50 atoms ) and corresponding high-precision DFT results (energy and force). For a well-trained model, given any corresponding configuration (include those not in the training data and supercells with a large number of atoms), its total energy and the force of each atom can be solved with minimal computational cost and at DFT-level accuracy. Thus, DP model can be applied as a force field in MD simulations.

In the training of a DP model for a certain material, one of the most important processes is to comprehensively sample various symmetrices structures. Considering the complex high-pressure environment of $\ce{LaH10}$, the candidate structure for the DP model training dataset need to be carefully determinate. We explored the low-enthalpy crystalline structures of the $\ce{La_{x}H_{y}}$ at pressures between 100 and 300 GPa via an evolutionary algorithm implemented in USPEX code\cite{Lyakhov2013,Oganov2011,Oganov2006}. The USPEX predicted structures are in good agreement with previous findings \cite{Liu2017} ( including  $Fm\bar{3}m$, $C2$ and $P\bar{1}$ phase of $\ce{LaH10}$ and other $\ce{La_{x}H_{y}}$ ). The enthalpies of the predicted structures are shown in Fig.~\ref{fig:method_part}(a). It can be observed that at pressures ranging from 170 to 300 GPa, the enthalpies of the $C2$, $P\bar{1}$, and $Fm\bar{3}m$ phases are lower compared to the other phases, indicating their greater stability within this pressure range. Therefore, the initial structure training set of the DP model are mainly from those three phases. Among the three structures, the $Fm\bar{3}m$ phase has higher symmetry, which will be referred to as the cubic phase later. The $C2$ phase and the $P\bar{1}$ phase are distorted version of the $Fm\bar{3}m$ phase, and they are referred to as low-symmetry phases later. Fig.~\ref{fig:method_part}(b) shows the schematic of cubic phase and low symmetry phase. Each $\ce{La}$ is surrounded by a cage structure composed of 32 $\ce{H}$ atoms, which contains 6 quadrilaterals and 12 hexagons. The simulation cells in this work obtain 44000 atoms to achieve statistical equilibrium. For more information about the DP and generation of the dataset, and parameters for DFT, please refer to the Supplemental Material\cite{Supplementary}.

\begin{table}[b]
\centering
\caption{Lattice constants of cubic phase at 0K predicted by DFT and DP model. The index of La-H and H-H bond represents in (1) and out(2) of the quadrilaterals in the cage structure.}
\label{lattice}
\begin{ruledtabular}
\begin{tabular}{ccc}
         & \textrm{DFT}   & \textrm{DP}    \\ \colrule
a/$\textrm{\AA}$        & 4.749 & 4.751 \\
$\alpha$/°              & 90    & 90    \\
La-H$^1$/$\textrm{\AA}$ & 1.9755 & 1.9764 \\
La-H$^2$/$\textrm{\AA}$ & 2.0564 & 2.0572 \\
H-H$^1$/$\textrm{\AA}$  & 1.1454 & 1.1453 \\
H-H$^2$/$\textrm{\AA}$  & 1.0645 & 1.0654 \\
La-La/$\textrm{\AA}$ & 3.3581 & 3.3595 \\
\end{tabular}
\end{ruledtabular}
\end{table}

\subsection{Quantum Thermal Bath}
DP potential is able to produce DFT-level energy and atomic force in MD simulation, and we use QTB to incorporated quantum effects in MD simulations. In MD classical limit, the equipartition theorem is fulfilled, therefore it only produce results that confirm to classical behavior\cite{Plimpton1995}. The core idea of QTB is based on the quantum mechanical fluctuation-dissipation theorem\cite{Callen1951}, it introduces associated random force and friction term into the equation, constituting a quantum thermal bath\cite{Dammak2009}. The equation of motion of a degree of freedom $x$ of a particle of mass $m$ in the presence of an external DP force $F(x)$ is modified to Langevin-like equation
\begin{equation}
m\ddot{x}=F(x)+\sqrt{2m\gamma}\Theta(t)-\gamma m\dot{x},
\end{equation}
where $\Theta(t)$ is a colored noise with a power spectral density
\begin{equation}
\Theta(\omega,T)=\hbar\omega[\frac{1}{2}+\frac{1}{exp(\frac{\hbar\omega}{k_BT})-1}],
\label{PSD}
\end{equation}
it includes the zero-point energy. 

Further, the numerical techniques related to QTB were improved\cite{Barrat2011}, and QTB can be easily manipulated and is independent of the studied system. Previous works have demonstrated that QTB can produce results in good agreement with experiments or PIMD results\cite{Wu2022,Dammak2009,Barrat2011,Dammak2011} with computation complexity comparable to classical MD.

\section{Results and Discussions}

\subsection{Validation of Deep Potential}

To obtain the DP model of $\ce{LaH10}$ used in this paper, 25427 candidate configurations which contain DFT energy and force information of $\ce{LaH10}$ were selected for the training dataset. All these configuration are generated by MD simulations, which adopt the isobaric-isothermal (NPT) ensemble with temperature set from 0 to 600 K, pressure set from 80 to 500 GPa. The comparison of the error between the DP model predictions and DFT results is shown in Fig.~\ref{fig:method_part} (c,d), the energy RMSE/Natoms and force RMSE are 3.272 meV/atom and 0.145 eV/\AA, respectively.

The lattice constants of the cubic phase predicted by DFT and DP model are shown in Table~\ref{lattice}, all constants including lattice parameters and angles predicted by DFT and DP are almost equal, indicating that our DP model is in good agreement with DFT calculations.

\begin{figure*}[htbp]
    \centering
    \includegraphics[width=0.88\textwidth]{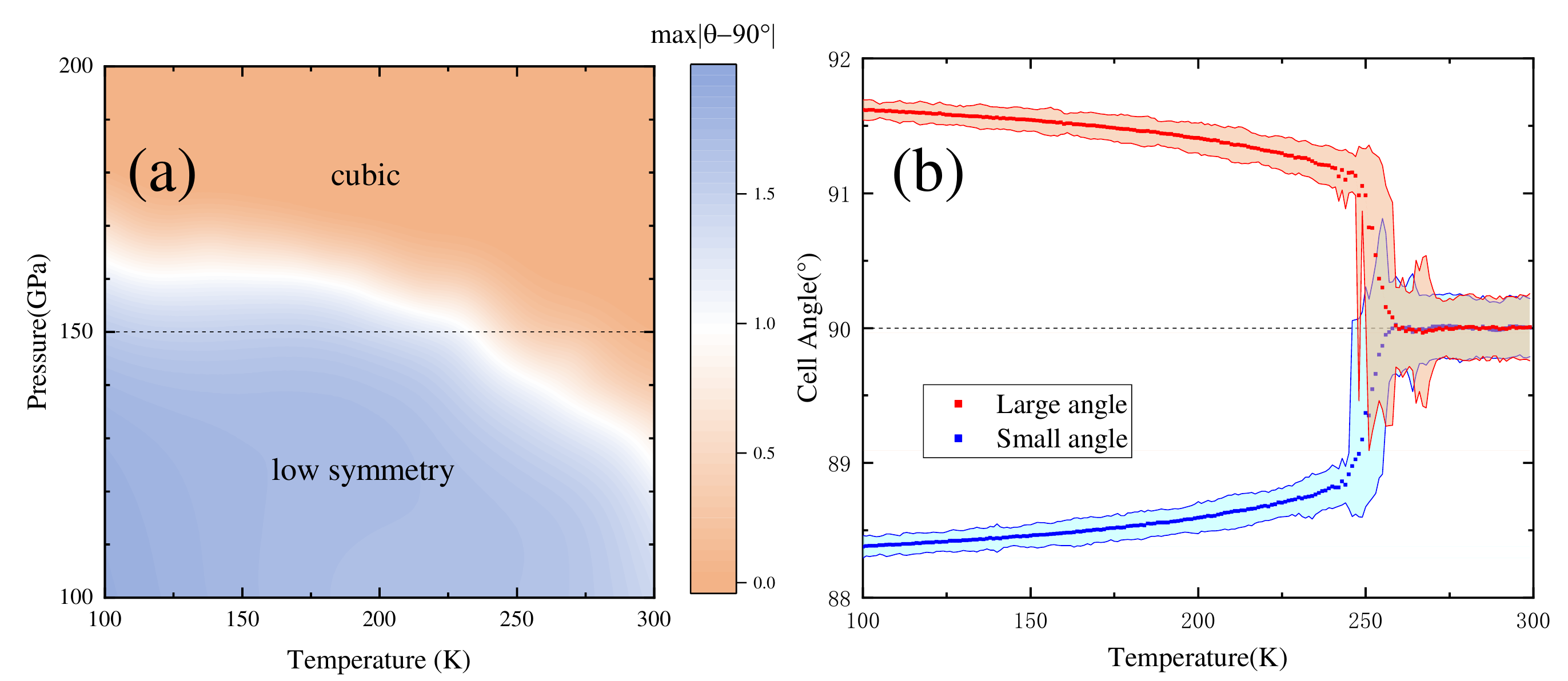}
    \caption{(a) Phase diagram of lattice structure of $\ce{LaH10}$ determined by classical MD, the colorbar is determined by the distortion angle $\theta$ compared to cubic phase. After incorporated quantum effects, the colorbar at this range all turn to 0. (b) The cell angle of H atoms quadrilateral as function of temperature under 150 GPa (dash line in (a)), red and blue dots represent the maximum and minimum cell angle, respectively.}
    \label{fig:structure}
\end{figure*}

\subsection{Pressure-Temperature Phase Diagram of $\ce{LaH10}$}
Using the DP model, the atomic structures of $\ce{LaH10}$ can be determined by MD simulations. During the MD simulations, the lattice constants and local structure of $\ce{LaH10}$ will dramatically change. So in order to intuitively distinguish the cubic phase and the low-symmetry phase from the perspective of atomic structure, we define the distortion angle $\theta$, which is the angle between the quadrilaterals in the cage structure and it is convenient for distinguish the structure phase transition. According to this definition, when the statistical results of $\theta$ is 90$^\circ$, the structure is cubic phase, otherwise it is low-symmetric phase.
In Fig.\ref{fig:structure}(a), we show the pressure–temperature phase diagram of $\ce{LaH10}$ obtained by the classical MD at first-principles accuracy. 
It is clear that cubic phase only stabilize at high pressure ($\ge$ 150 GPa), which is in good agreement with experimental results\cite{Drozdov2019}. In addition to the effect of pressure, the temperature effect also affects the stability of the cubic phase. When the temperature increases, the stable pressure of the cubic phase will decrease, which is consistent with the description of thermal effects predicted by other theoretical result\cite{Watanabe2022}. 
Another point can be concluded is, the transition between cubic and low-symmetry phase is first order transition. To further explore this transition, we show the cell angle of H atoms quadrilateral as function of temperature under 150 GPa (dash line in (a)) in  Fig.\ref{fig:structure}(b). We found that whether the angle is greater or less 90°, they all change steadily with the increase of temperature before 250K, and the angle rapidly approaches 90° near 250K. We can also see white narrow line in Fig.\ref{fig:structure}(a) between these two phases, it means for any pressure, this transition is rapid. 

\begin{figure}[htbp]
    \centering
    \includegraphics[width=0.88\columnwidth]{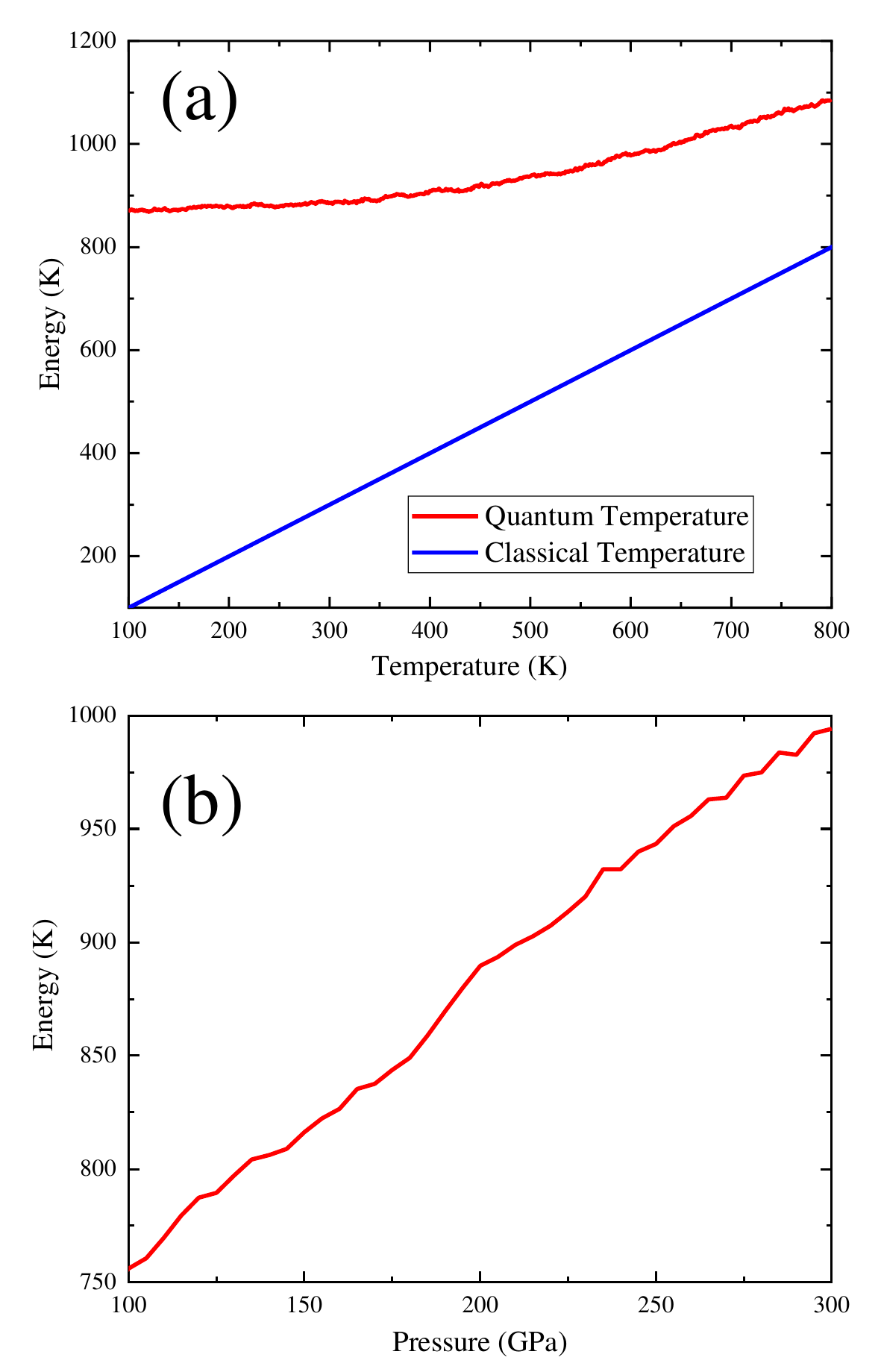}
	 \caption{(a) Comparison of the average energy per atom in $\ce{LaH10}$ as a function of temperature between classical and quantum simulations. (b) Zero-point energy of $\ce{LaH10}$ as a function of pressure.}
    \label{quantum_temperature}
\end{figure}

\begin{figure*}[htbp]
    \centering
    \includegraphics[width=0.88\textwidth]{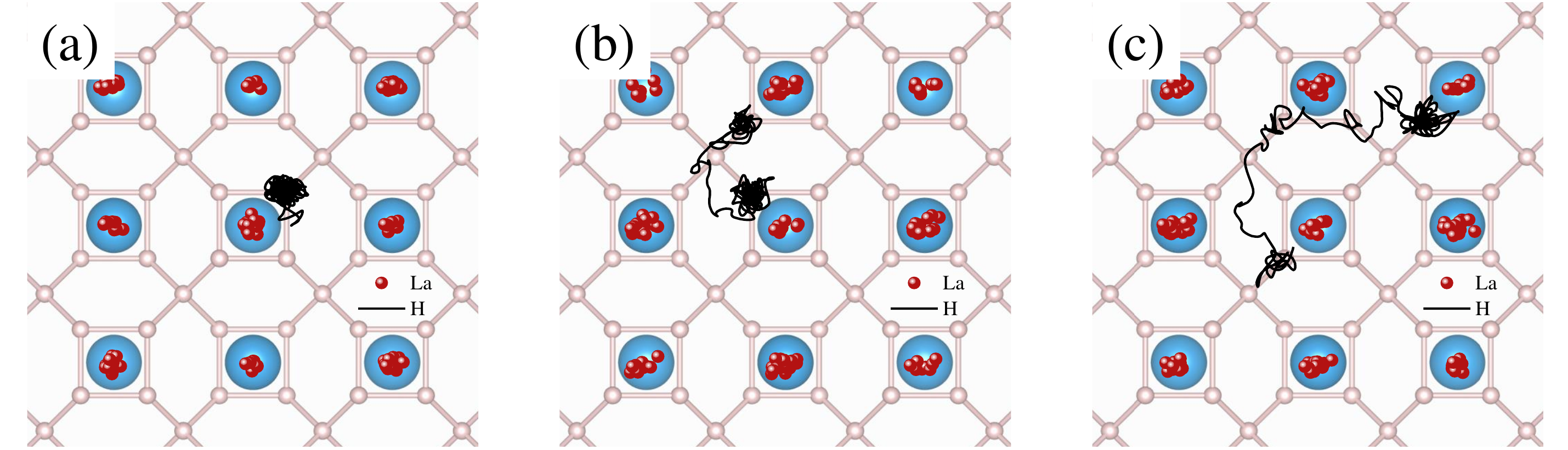}
    \caption{DP+QTB simulations under 200 K and 200 GPa, three of typical trajectories of $H$ atoms and local $La$ atoms: (a) no exchange (b) one exchange and (c) multiple exchanges.}
    \label{quantum_diffusion}
\end{figure*}

While in the simulation combined with QTB, as long as pressure is above 100 GPa, $\ce{LaH10}$ will turn into cubic phase across the entire temperature range, the corresponding phase diagram is shown in Supplemental Material\cite{Supplementary}. This again proves that the cubic phase is more stable than other phases under high pressure, and quantum effects play key a role in stabilizing it, which is consistent with other works\cite{Errea2020}. It is due to the nuclear quantum effect of $\ce{LaH10}$ is significant, which enhances the motion of atoms and makes it have an equivalent thermal effect, and making the cubic phase stable. To characterize the strength of the quantum effect at any temperature, we compute the evolution of energies (classical and quantum) as a function of temperature and show it in the form of $\frac{E}{3Nk_B}$, as shown in Fig.~\ref{quantum_temperature}(a). The energy of $\ce{LaH10}$ in the classical case are shown in blue line, they are equal to $3Nk_BT$ at any temperature as dictated by the equipartition theorem of classical Boltzmann statistics. The DP+QTB simulation results are shown in red line. At 0K, the energy of $\ce{LaH10}$ is not 0, which is the zero-point energy. The zero-point energy of $\ce{LaH10}$ is as high as $\sim$900 K, which indicates the nuclear quantum effects of $\ce{LaH10}$ is significant. And this is consistent to the simple harmonic approximation, effective spring coefficient of $\ce{LaH10}$ is very high due to the high pressure, and mass of hydrogen is low. To further measure the strength of the quantum effect, we use the definition of zero-point temperature $T_{zero}=\frac{1}{k_B}\int_0^\infty g(\omega)\frac{1}{2}\hbar\omega d\omega$, it represents the average energy per atom of zero-point fluctuation in the form of temperature. The variation of the zero-point energy of $\ce{LaH10}$ with pressure is illustrated in Fig.~\ref{quantum_temperature}(b). As the pressure increases, the zero-point energy also increases, suggesting that pressure amplifies the zero-point motion of atoms. The underlying cause of this phenomenon is attributed to higher pressure, which results in an elevated effective spring coefficient.

\subsection{Quantum Diffusion of Proton}

Since the quantum effects have such great influence on the energy and atomic motion, it naturally affects atomic displacement, especially for the light element H\cite{Wang2021}. We find that in our DP+QTB simulation results, the motion of H atoms is much more intense than that of La atoms. As shown in Fig.~\ref{quantum_diffusion}, we show three typical types of trajectory of H and La atoms under 200 K and 200 GPa. In all three types the La atoms vibrate near its equilibrium lattice position, most protons keep vibrating near equilibrium lattice position (Fig.~\ref{quantum_diffusion} (a)), part of protons diffuse away from the initial neighbor La atoms, and two protons will finally exchange their  location(Fig.~\ref{quantum_diffusion} (b,c)).

\begin{figure}[b]
    \centering
    \includegraphics[width=\columnwidth]{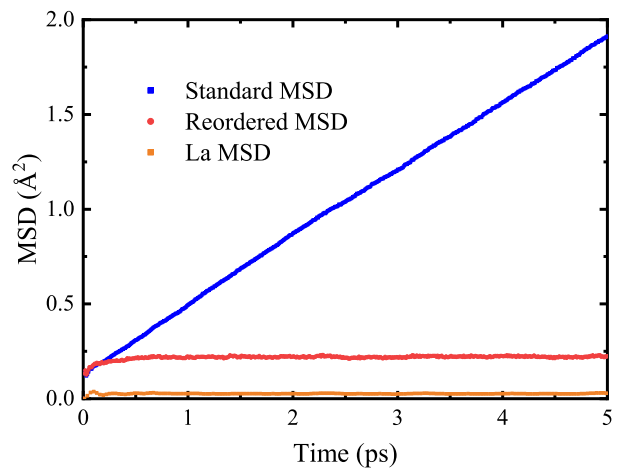}
    \caption{Standard MSD and reordered MSD of proton and MSD of La atoms in $\ce{LaH10}$ as a function of time simulated by DP+QTB under 5 K 200 GPa.}
    \label{msd}
\end{figure}

It is worth emphasizing that this proton diffusion phenomenon is not induced by thermal fluctuation\cite{Liu2018}, but induced by the quantum effects of zero point fluctuation, thus the proton diffusion can also appear under low temperature without thermal effects. The mean square displacement (MSD) curves of the protons under 5 K and 200 GPa are shown in Fig.~\ref{msd}. We recorded the average MSD of each H atom as a function of simulation time, as shown by the blue dots in the Fig.~\ref{msd}. The MSD of protons is increase with the time, while in solid such as crystal it is a constant that does not change over time. And its diffusion coefficient obtained from this curve is $3.5 \times 10^{-5} cm^2/s$, which is remarkably the same order of magnitude as the liqui\cite{Mills1973}. These all indicate that the behavior of $\ce{LaH10}$ is different from normal solids. Also, since the atom La is heavy, naturally its MSD is negligible, and does not increase over time, as shown in Fig.~\ref{msd}. This can proves that even part of $\ce{LaH10}$ is mobile, the framework is still rigid, it will remain solid.

Considering the indistinguishability of microscopic particles, it is difficult to directly distinguish the diffusion and exchange of H atoms through existing experimental observation methods. According to the standard MSD definition, the simulation results obtained by tracking the diffusion of each atom may be overestimated compared to experimental results. Therefore, we define a reordered MSD (MSD$'$) that no longer uses atomic order as the basis for tracking, 
\begin{equation}
MSD'=\frac{1}{N}\underset{\mathcal{P}}{\min}\sum_i{\left| \boldsymbol{r}^{\mathcal{P}\left( i \right)}\left( t \right) -\boldsymbol{r}^i\left( 0 \right) \right|}^2,
\end{equation}
where $\mathcal{P}$ is the permutations of atom order.
By this definition, no matter protons diffuse to any perfect lattice location and vibrate near their new equilibrium location (for example the situations of Fig.~\ref{quantum_diffusion} (b) and (c)), the value of MSD$'$ will be consistent with the oscillation near the equilibrium position. The MSD$'$ results are shown by red dots in Fig.~\ref{msd}, compared to standard MSD, their magnitude is much smaller and almost a constant over time. This suggests that certain experimental characterization methods may not be able to detect the diffusion of protons in $\ce{LaH10}$. Alternative techniques, such as isotope labeling methods, may provide better insight into this phenomenon.

\begin{figure}[htbp]
    \centering
    \includegraphics[width=\columnwidth]{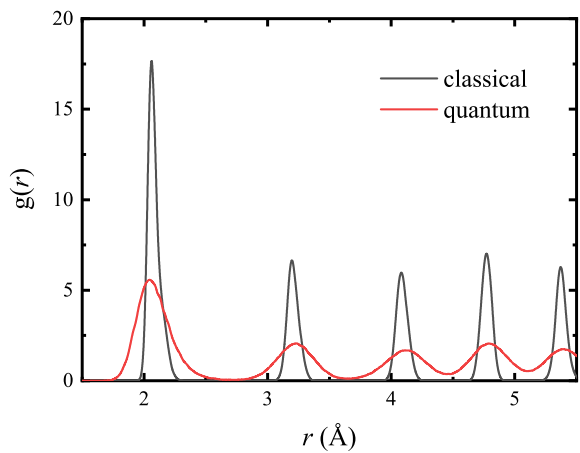}
    \caption{Partial radial distribution function (La-H) of $\ce{LaH10}$ under classical and quantum situations (1 K).}
    \label{RDF}
\end{figure}

High mobility of protons is likely to affect the structural stability of $\ce{LaH10}$, however the structure of $\ce{LaH10}$ should remain solid. To find out this problem, in Fig.~\ref{RDF}, we show the partial radial distribution function (pRDF) between La and H atoms in both classical and quantum situations, it can represent the local environment of $\ce{LaH10}$. The locations of several peaks of the classical and quantum pRDFs have a very high degree of coincidence, which indicates that in the quantum situation the distribution of H relative to La is consistent with the classical situation. Considering that we mentioned earlier that La has a small degree of movement and stays near the perfect lattice position. So in $\ce{LaH10}$ even though protons are highly mobile, La atoms will keep their rigid framework, thus $\ce{LaH10}$ still maintains a relatively stable structure. In addition, the peak broadening of pRDF is larger in the quantum situation, which is caused by the fact that the atoms exhibit zero-point vibrations in the quantum situation, and their spatial position distribution is more delocalized.

\subsection{Discussion}

The zero-point motions of protons in $\ce{LaH10}$ are compared to the near-neighbor distance, and it aligns with the definition of quantum crystals\cite{Nosanow1966,Guyer1970}, similar behavior can be found in solid He. Now we know that part of mass of $\ce{LaH10}$ is able to flow due to quantum effects, but its structure is still rigid, which is a remarkable consequence can be found in supersolid\cite{Balibar2010}. Supersolidity was previously only reported in pure H or He systems, because the zero-point motion of these two systems are obviously intense. Since the zero-point motion of protons in $\ce{LaH10}$ is intense, perhaps hydrogen-rich supercondutors can also become a candidate material system for supersolid research. Moreover, this protons diffusion phenomena may also have impacts on superconductivity of $\ce{LaH10}$, such high T$_{c}$ may related to protons diffusion.

\section{CONCLUSION}

We develop a deep-learning based DP model to describe the energy and atomic forces of hydrogen rich superconductor $\ce{LaH10}$, which has DFT level accuracy. Using this model for classical MD calculations, we confirm that for $\ce{LaH10}$, superconductivity-favored cubic phase is more stable above 150 GPa, which is consistent with other theoretical and experimental results. Using DP+QTB method, we successfully incorporate quantum mechanics in MD simulation of $\ce{LaH10}$. The strength of quantum effect of $\ce{LaH10}$ is proven to be significant, and pressure is able to regulate the strength of the quantum effect. Such significant quantum effect results in strong zero-pint motion of the protons, which allows them to diffuse over the lattice, but the structure of $\ce{LaH10}$ can still remain rigid. This partially intense diffuse phenomenon may named as quantum crystal like behavior, and may provide some new insights for the study of the structure and superconductivity of hydrogen-rich superconductors.

\begin{acknowledgments}
This work was supported by the National Key R$\&$D Program of China (Grants No. 2021YFA0718900 and No. 2022YFA1403000), the Key Research Program of Frontier Sciences of CAS (Grant No. ZDBS-LY-SLH008), the National Nature Science Foundation of China (Grants No. 11974365 and No. 12204496), the K.C. Wong Education Foundation (GJTD-2020-11), and the Science Center of the National Science Foundation of China (52088101). 
\end{acknowledgments}

\end{document}